*Full Length Research Paper*

# Development finance institutions (DFIs), political conditions, and foreign direct investment (FDI) in Sub-Saharan Africa

Carmen Berta C. De Saituma Cagiza[1]* and Ilidio Cagiza[2]

[1]School of Social Science and Global Studies, College of Arts and Sciences, University of Southern Mississippi, Mississippi, USA.
[2]Imperial Business School, Imperial College, London, UK.

This study investigates the dynamic relationship between development finance institutions (DFIs), foreign direct investment (FDI), and economic development in Sub-Saharan Africa (SSA) from 1990 to 2018, using a quantitative panel dataset of annual data for five SSA countries (Nigeria, Ghana, Kenya, South Africa, and Zimbabwe) and a fixed-effects model estimated in STATA. Specifically, the analysis examines whether DFIs enhance FDI inflows, thereby promoting economic growth and contributing to the achievement of the Sustainable Development Goals (SDGs). The findings indicate that although DFIs have a theoretically positive impact on FDI, this relationship is not statistically significant across the sample, suggesting contextual dependencies influenced by regional economic variations. The study also analyzes how economic growth, trade openness, inflation, political stability, and the rule of law influence this nexus, elucidating their roles in shaping investment climates. A sectoral analysis indicates that DFI investments in infrastructure, agribusiness, and finance significantly affect FDI, with infrastructure having the greatest impact owing to its foundational role in economic systems. This research contributes by linking DFIs with FDI in SSA in a panel setting, thus providing a framework for policymakers to strengthen institutional and macroeconomic conditions to optimize the impact of DFIs on FDI and, ultimately, on sustainable development. The findings underscore the need for targeted policies to address regional disparities and enhance DFI effectiveness in fostering sustainable growth.

## INTRODUCTION

This paper investigates how development finance institutions (DFIs), foreign direct investment (FDI), and broader economic trends intersect within Sub-Saharan Africa (SSA). Particular attention is given to the extent to which these mechanisms support national growth efforts and contribute to long-term development goals, including

*Corresponding author. E-mail: carmen.cagiza@allusainvestment.holdings. Tel: +1 832 484 8101.

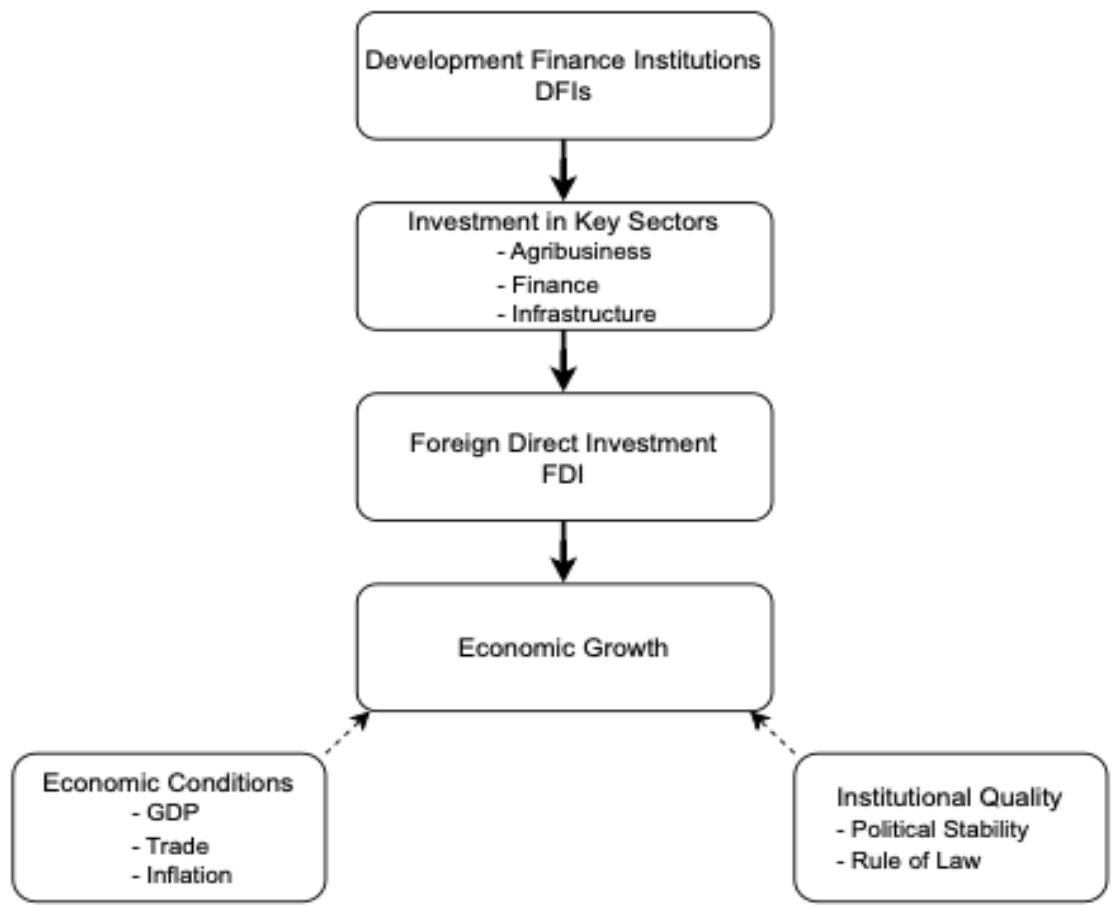

**Figure 1.** Conceptual framework of DFIs, FDI, and economic growth in SSA.

those outlined in the Sustainable Development Goals (SDGs). The study is guided by a central research question: do DFIs influence the volume and effectiveness of FDI in SSA, and does this, in turn, translate into measurable economic progress? Additionally, the analysis considers how institutional strength and macroeconomic conditions shape these interactions. This study addresses a notable gap in the literature by examining DFIs' impact on FDI in SSA, an aspect few studies have explicitly linked, and by incorporating sector-specific DFI data and institutional factors into the analysis. This focus allows for an evaluation of the unique contribution of DFIs to FDI inflows in the region.

Figure 1 illustrates the hypothesized relationships, depicting how DFIs drive investment in key sectors, agribusiness, finance, and infrastructure, thereby influencing FDI and, ultimately, economic growth in SSA. Mediating factors, including economic conditions (GDP per capita, trade openness, inflation) and institutional quality (political stability, rule of law), are represented by dashed arrows, indicating their moderating effects on the pathway from FDI to economic growth. It is hypothesized that DFIs play a pivotal role in this context due to their capacity to support the private sector, intervene in challenging environments, and mobilize financial resources (Attridge and Engen, 2019; Massa, 2011). These institutions catalyze economic growth by strengthening local initiatives in sectors vital for development, such as agribusiness, finance, and infrastructure. However, despite their potential, the effectiveness of DFIs in SSA has been less than satisfactory, with the region lagging in economic growth relative to global standards (Bodomo, 2017; Bräutigam, 2011). DFIs are also often "first movers" in fragile markets, absorbing initial investment risks and demonstrating viability to attract other investors (Collier and Venables, 2021). By examining DFIs alongside FDI, this study evaluates whether DFIs have lived up to their promise of spurring additional private investment in SSA. To address these disparities, the analysis includes several key variables that influence FDI inflows in SSA:

1. Market size and economic development (proxied by GDP per capita): Larger or more developed economies are generally more attractive to foreign investors due to higher demand and better infrastructure. For instance, Abdi et al. (2024) find that GDP (market size) is a significant driver of FDI in African countries. Similarly, robust GDP per capita rates can signal market potential and investment opportunities (Keeley and Matsumoto, 2018), although this study primarily uses GDP level rather than growth rate.

2. Trade openness: Defined as the sum of exports and imports as a share of GDP, trade openness positively influences FDI by facilitating market access and promoting economic integration (World Bank, 2018).

3. Inflation rates: The relationship between inflation and FDI is nuanced. Inflation is included to capture macroeconomic stability; high inflation typically signals uncertainty and increased costs, potentially deterring FDI, whereas moderate inflation might coincide with rapid growth (Kore, 2019).

4. Political stability: Measured using the World Bank's Political Stability Index, political stability can reduce perceived investment risks and thus support FDI inflows (World Bank, 2018)..

5. Rule of law: Reflects the strength of legal institutions and property rights. A stronger rule of law provides investors with confidence in contract enforcement and property protection, thereby encouraging FDI (Acemoglu et al., 2001; North and Thomas, 1973).

Recognizing the diverse economic needs within SSA, the analysis extends to a sectoral examination of DFI investments, focusing on the following:

1. Agribusiness: Investments in this sector are vital for food security, rural development, and employment generation, potentially making the industry attractive for FDI.

2. Finance: DFI investments that enhance financial inclusion and develop financial markets can fundamentally alter the investment landscape by reducing the cost of capital and risk.

3. Infrastructure: Investment in infrastructure (that is, transport, energy, water, information and communications technology) is essential for diversifying the economy and improving the business environment, acting as a significant pull factor for FDI.

This study fills a gap in the literature by investigating the broad impact of DFIs on FDI and providing a nuanced understanding by including these critical variables and conducting a sector-specific analysis. While previous studies have focused on the microlevel effects of DFIs or have broadly discussed FDI in developing countries, there is a lack of comprehensive research specifically linking DFIs, FDI, and economic growth in SSA, while considering the role of financial and institutional factors. Previous studies on FDI in Africa have largely examined determinants or outcomes other than the role of DFIs. For instance, Cleeve (2008) analyzed fiscal incentives to attract FDI, Magombeyi and Odhiambo (2017) investigated the link between FDI and poverty reduction, and Kore (2019) focused on macroeconomic determinants of FDI in SSA. However, few works have directly assessed the impact of DFIs on FDI. While Pietro (2013) discussed the potential of African DFIs and Attridge and Engen (2019) highlighted DFIs' ability to mitigate investment risk in high-risk environments, empirical analyses explicitly linking DFI activities to FDI inflows in SSA have been lacking. This study addresses this gap by providing a detailed panel analysis from 1990 to 2018, using data from five SSA countries (Nigeria, Ghana, Kenya, South Africa, and Zimbabwe) selected for their economic diversity and data availability. The period 1990–2018 is chosen because it captures the era of economic liberalization and FDI surge in SSA post-1990, while ending before the major disruptions of 2019–2020; this ensures data consistency and allows for observing nearly three decades of trends during which DFIs became increasingly active in the region.

The timing of this study is pertinent. As SSA aims to accelerate investment-led growth in the wake of recent global shocks, DFIs are increasingly being called upon to mobilize private capital for development (Attridge and Engen, 2019). Meanwhile, FDI flows to Africa have been volatile but showed a strong rebound in 2021 (UNCTAD, 2022), underscoring the importance of understanding how DFI efforts and domestic conditions interact to influence investment. In summary, this paper contributes to the development finance literature by linking DFIs with FDI and economic growth outcomes in SSA. It provides evidence on whether DFIs have the intended catalytic effect on private investment and highlights the importance of supportive economic and institutional conditions. However, certain limitations are acknowledged: for example, the country sample is not random and is limited to five nations, and there are potential endogeneity issues (such as reverse causality between FDI and growth or non-random DFI allocation) that are not fully resolved in the current framework. Future research directions are suggested to address these issues.

## LITERATURE REVIEW AND HYPOTHESIS DEVELOPMENT

### FDI, growth and development

Prior studies have extensively explored FDI's role in promoting development in emerging economies. FDI is typically defined as an investment by a company or individual in productive assets of another country (Maurice, 2009). It has been promoted by governments as a tool to stimulate job creation, technology transfer, and trade competitiveness (Keeley and Matsumoto, 2018). Empirical findings on FDI's development impact are mixed; positive effects such as productivity gains and poverty reduction are often noted, but outcomes vary widely depending on host-country conditions. For instance, FDI's contribution to growth may be conditioned on factors like human capital, financial market development, or institutional quality (Baiashvili and Gattini, 2020). Some studies caution that the link between FDI and long-term development is not automatic or linear (Ahmeti and Kukaj, 2016; Baiashvili and Gattini, 2020). The benefits of FDI can be fully realized only when host countries have a conducive environment to absorb and retain spillovers (e.g., stable policies, educated workforce, sound institutions). Due to distinct economic conditions across countries, no single host environment can uniformly attract all types of FDI. Investment decisions are often made based on investor-specific criteria, including policy stability, resource availability, and perceived profitability (Cuyvers et al., 2011). Recent evidence suggests that the FDI-development relationship is highly context-dependent, varying with local conditions (Baiashvili and Gattini, 2020). In SSA, where conditions vary greatly, the effects of FDI on growth have been found to be highly context-dependent (Magombeyi and Odhiambo, 2017).

**Institutions and investment climate**

The institutional framework is central to understanding economic development, particularly in emerging economies. Institutions, comprising formal laws and informal norms, shape investment climates by influencing the transaction costs and enforcement mechanisms. Economic institutions influence multiple aspects of development, including access to and distribution of income, assets, and knowledge (Acemoglu and Robinson, 2012). These institutions not only affect macroeconomic performance but also shape the broader allocation of resources in ways that impact equity and long-term stability. Institutions that protect property rights, facilitate investment, and ensure accountability, such as those discussed by Acemoglu et al. (2001) and Tchouassi (2014), are often linked with sustained development outcomes. Conversely, weak institutions can deter FDI or limit its benefits.

While the literature on DFIs remains relatively limited, scholars increasingly recognize their importance in the broader development finance framework. There is growing acknowledgment that DFIs, when well-structured, can help attract foreign investment, especially in contexts where traditional markets perceive high levels of risk (Attridge and Engen, 2019). Financial systems in many developing countries often face challenges in attracting capital due to risk perceptions and information gaps. To address these challenges, financial institutions generate informational efficiencies by mitigating transaction costs and reducing asymmetries in the flow of capital (Carbaugh, 2015; Rodrik, 2007).

**Role of DFIs**

DFIs are specialized bodies (often government-backed or multilateral) that provide financing and support to projects in developing regions. They aim to bridge market gaps by investing in several sectors or environments that private investors might shun due to high risk or lower immediate returns. DFIs, when well-structured, can help attract foreign investment by improving risk-return profiles, especially in high-risk contexts (Attridge and Engen, 2019). In Africa, a significant share of blended finance and impact investment is channeled through DFIs, and recent estimates suggest DFI investments account for roughly 15-20% of total FDI flows to Africa (Florian, 2024). DFIs offer long-term loans, equity, guarantees, and technical assistance to reduce investment risks (De Luna-Martinez and Vicente, 2012). By doing so, DFIs create demonstration effects, showing that viable investments are possible in markets that were previously seen as too risky. DFIs also help mobilize domestic savings, spur private sector investment, and foster innovation-led growth. Although theoretical perspectives differ, recent analyses continue to support the crucial role of financial institutions in development. Collier and Venables (2021) provide an insightful discussion on how DFIs can pioneer investments in fragile states and crowd-in private capital by bearing the initial high risks. The analysis is predicated on the idea that if DFIs are effective, a positive association between DFI activity and FDI inflows should be observed, after controlling for other factors. Conversely, if no such association is found, it might imply that DFIs are not sufficiently catalytic or that other conditions need to be in place.

**Hypotheses development**

Drawing on the above literature, the following hypotheses are formulated about the expected relationships in this study (each hypothesis corresponds to one of the key variables in the model):

H1 (DFIs and FDI): DFIs are positively associated with increased FDI inflows in SSA (Kore, 2019; Massa, 2011). In other words, countries and years with higher DFI investment should, on average, see higher FDI, consistent with a catalytic role of DFIs.

H2 (Economic development – GDP): Higher levels of economic development and market size (approximated by GDP per capita) are associated with greater FDI inflows (Keeley and Matsumoto, 2018). It is expected that wealthier or larger economies will attract more FDI (Abdi et al., 2024). This hypothesis aligns with the notion that market-seeking investors favor countries with larger consumer bases and better infrastructure.

H3 (Inflation): Higher inflation rates deter FDI inflows, as inflation is a sign of macroeconomic instability and can erode investment returns (Kore, 2019). Thus, an inverse relationship between inflation and FDI is hypothesized.

H4 (Trade openness): Greater trade openness correlates with increased FDI, since an open economy with high trade (imports + exports relative to GDP) offers foreign investors better access to international markets and inputs (World Bank, 2018).

H5 (Political stability): Stable political environments contribute positively to FDI attractiveness (Sabir et al., 2019; World Bank, 2018).

H6 (Rule of law): A strong rule of law (effective legal frameworks, property rights, contract enforcement) positively influences FDI by providing legal certainty and protecting investors' rights (Acemoglu et al., 2001; North and Thomas, 1973).

These hypotheses form the basis of the empirical analysis and conceptual framework.

## METHODOLOGY

### Data sources and sample

This study employs a quantitative panel data analysis to explore how DFIs influence FDI inflows and economic development in SSA over the period 1990 to 2018. The longitudinal design allows examination of temporal dynamics over this 29-year period. The sample comprises five SSA countries: Nigeria, Ghana, Kenya, South Africa, and Zimbabwe. These countries were selected based on data availability and economic diversity, together they represent a mix of large and medium economies, various regions of SSA, and differing degrees of DFI and FDI engagement. The sample is non-random, constrained by the availability of consistent data across all variables. However, the selection captures a broad range of SSA contexts (from the relatively diversified economy of South Africa to resource-rich Nigeria and more agrarian economies like Kenya or Ghana), providing valuable insights. A limitation is that results may not generalize to all SSA countries.

### Variables and measurement

The dependent variable is FDI Inflows, measured as annual net FDI inflows into the country as a percentage of GDP (World Bank, 2018). This standard measure captures FDI relative to the size of the economy. The key independent variables are:

1. DFI commitments: The value of multilateral DFI investment commitments (new projects or funding agreements) in each country-year, expressed as a percentage of that country's GDP. This was constructed from the World Bank and major multilateral DFI project databases, focusing on commitments in three sectors of interest: agriculture (AGRI), finance (FIN), and infrastructure (INFRA). Each year's DFI figure represents the sum of all new commitments signed in that year (as opposed to disbursements). For example, if a DFI approved a $50 million infrastructure project in Country X in 2010 and $30 million in 2011, the data record 2010: DFI = $50m, 2011: DFI = $30m (each normalized as % of GDP). The methodology follows the practice in Massa (2011) of considering commitment values in analysis. By expressing commitments as a share of GDP, the analysis controls for country size and facilitates comparison across countries and years.
2. GDP per capita: Gross domestic product per capita (constant US$, base year consistent across the period) is used as a proxy for economic development and market size. In the analysis, this variable is labeled GDP for brevity. Higher GDP per capita is expected to attract more FDI (larger market with higher purchasing power). GDP growth rates were also computed for reference, but the regression mainly includes the level measure to avoid multicollinearity between growth and level (the two are correlated in the data).
3. Trade openness (TRD): Calculated as the ratio of the sum of exports and imports to GDP. Data are from World Bank (2018).
4. Inflation (INFL): Annual percentage change in consumer prices (CPI), sourced from World Bank (2018). It is used as an indicator of macroeconomic stability.
5. Political Stability (Pol): Index of Political Stability and Absence of Violence/Terrorism from the World Bank's Worldwide Governance Indicators. Originally, this index ranges approximately from –2.5 (very unstable) to 2.5 (very stable).
6. Rule of law (Law): Index from the Worldwide Governance Indicators reflecting perceptions of the extent to which agents have confidence in and abide by the rules of society (property rights, law enforcement, courts).

### Sectoral analysis of DFI investments

To examine how DFIs operate across different areas of the economy, a sector-level analysis was conducted. This allowed for identification of which sectors attracted the highest levels of foreign capital and contributed most significantly to economic expansion in SSA. Three variables were created: DFI_AGRI, DFI_FIN, and DFI_INFRA, which represent the annual DFI commitments (as % of GDP) in agribusiness, financial sector, and infrastructure projects respectively. These come from disaggregating the total DFI commitments using project-level data. Not every country-year has DFI activity in all three sectors; where no project was recorded in a sector for a given year, the value is zero.

### Econometric controls and estimation approach

Given the observational nature of this study, experimental or matched-group designs were not appropriate. To address unobserved heterogeneity across countries, a fixed-effects estimation method was employed, which accounts for characteristics that do not change over time. Year dummies were included in the model to absorb any time-specific shocks or common temporal trends that might affect the dependent variables. Data were collected annually over the study period, ensuring consistency across variables. Any discrepancies in the data were resolved by cross-referencing with the primary sources or official publications. Although the fixed-effects approach controls for time-invariant differences across countries, potential endogeneity concerns are acknowledged in the model. For instance, there could be reverse causality (e.g., higher FDI inflows might attract more DFI projects) or omitted variable specificity affecting the results. No specific remedies for endogeneity (such as instrumental variables or a dynamic panel estimator) were implemented due to data and scope limitations.

### Model specification and hypothesis

The baseline panel model estimated is:

$$FDI_{it} = \beta_0 + \beta_1 DFI_{it} + \beta_2 GDP_{it} + \beta_3 INFL_{it} + \beta_4 TRD_{it} + \beta_5 POL_{it} + \beta_6 Law_{it} + \alpha_i + \delta_t + \varepsilon_{it}$$

where it = $i$ indexes country and $t$ indexes year; αi accounts for country-specific fixed effects; δt captures time-related fixed effects; εit is the error term. This specification tests H1–H6 jointly. For the sectoral analysis, the extend the model by replacing the aggregate DFI term with the three sector-specific DFI variables:

$$FDI_{it} = \beta_0 + \beta_1 DFI\_AGRI_{it} + \beta_2 DFI\_FIN_{it} + \beta_3 DFI\_INFRA_{it} + \beta_4 GDP_{it} + \beta_5 INFL_{it} + \beta_6 TRDL_{it} + \beta_7 Pol_{it} + \beta_8 Law_{it} + \alpha_i + \delta_t + \varepsilon_{it}$$

Including DFI_AGRI, DFI_FIN, and DFI_INFRA in one model allows us to see which sector's DFI commitments have the strongest association with FDI while controlling for the others. This combined approach assumes that sectoral DFI effects are additive and independent (a separate regressions was ran for each sector's DFI in case of multicollinearity, and results were qualitatively similar, with infrastructure DFI consistently showing the largest effect).

**Hypotheses**

H1: DFIs have a positive effect on FDI inflows in SSA
H2: Higher GDP per capita is associated with increased FDI inflows
H3: High inflation rates deter FDI inflows
H4: Greater trade openness is associated with higher FDI inflows
H5: Greater political stability is associated with increased FDI inflows
H6: A stronger rule of law is associated with increased FDI inflows

These hypotheses correspond to the signs of coefficients $\beta_0$ through $\beta_6$ in the panel model earlier mentioned.

**Statistical analysis**

All statistical computations were performed using STATA software. To improve data accuracy, entries were verified against multiple authoritative sources, including databases from the United Nations and the World Bank (United Nations Conference on Trade and Development [UNCTAD], 2022; World Bank, 2018). Any discrepancies were resolved through triangulation with official datasets or cross-referenced publications.

Automated diagnostics within STATA were used to assess data quality and identify potential outliers or gaps. Standard robustness checks were applied, including normality assessments (e.g., Jarque-Bera test), multicollinearity tests, and heteroskedasticity tests (e.g., White test) to validate model assumptions. The STATA code was independently reviewed to ensure correctness. Sensitivity checks were conducted to evaluate whether the results held when selected variables were added or removed from the model. The panel included 145 observations, though some data gaps existed due to missing values in specific years or countries. This issue was addressed by focusing on consistently available data to maintain analytical integrity.

Country selection, Nigeria, Ghana, Kenya, South Africa, and Zimbabwe, was informed by several factors: availability of long-term data, economic diversity, sectoral relevance, and engagement with DFIs. Countries with substantial data limitations or inconsistent reporting across key variables were excluded from the final sample to maintain representativeness.

**RESULTS AND DISCUSSION**

**Diagnostic tests**

To assess whether the dataset met the assumptions required for valid inference, several diagnostic checks were conducted. First, normality was tested using the Jarque-Bera method, which evaluates whether the residuals are consistent with a normal distribution (Greene, 2012). The computed Chi-square statistic for the test was 0.081, suggesting that the null hypothesis of normally distributed residuals cannot be rejected, indicating that the residuals conform to expected distributional assumptions.

Next, the model was evaluated for multicollinearity using variance inflation factors (VIF). Multicollinearity is a concern when two or more predictors are highly correlated, which can distort coefficient estimates and weaken the precision of inference (Wooldridge et al., 2017). As shown in Table 1, all VIF values fall well below the commonly used threshold of 10, indicating that multicollinearity is not a significant issue in this dataset. The VIF scores in Table 1 confirm that the independent variables are not highly intercorrelated, with the highest recorded VIF being 1.15, far below the level at which variable redundancy would typically raise concern. These results validate the reliability of the individual predictors in the regression model.

Table 2 provides the correlation matrix for all variables included in the study. The pairwise correlations between explanatory variables are low to moderate, and none exhibit strong multicollinearity. The positive associations observed between FDI and most predictors, particularly GDP per capita and trade openness, are consistent with the study's theoretical expectations. These findings further support the assumption that DFIs and macro-institutional indicators meaningfully contribute to variations in FDI inflows across SSA.

**Descriptive statistics**

Table 3 presents descriptive statistics for the primary model variables. FDI as a share of GDP has a mean of about 5.3% with a standard deviation of 9.1%. The minimum FDI value is –3.73% of GDP, indicating an instance of net disinvestment, while the maximum is 72.07%, likely due to a singular large investment relative to a small economy.

DFI commitments average about 30.5% of GDP with a standard deviation of 16.96%. However, this requires careful interpretation: since DFI commitments are expressed as % of GDP, large project commitments in small economies or recession years can cause high ratios. The minimum non-zero DFI/GDP is 0.32%, and the maximum is 74.62%, indicating that in one country-year, new DFI commitments amounted to about 0.75 times the country's GDP. On average, DFI commitments are a few percentage points of GDP each year for most observations, with the mean inflated by a few big spikes.

The GDP per capita variable has a mean of about $1,015 and a standard deviation of $1,719, reflecting the wide disparity in development levels among the sample countries. The lowest observed GDP per capita is $85.55, while the highest is $8,605, underscoring the heterogeneity in the panel.

Trade openness has a mean of 0.7786 (approximately 77.9% of GDP). An average of 0.78 indicates that trade flows are roughly 78% of GDP on average. Notably, the maximum value appears to be inconsistent; many SSA countries in the sample have trade/GDP ratios around 50-70%, with some years exceeding 100% in smaller economies. Any inconsistencies in the dataset were

Table 1. Multicollinearity test using variance inflation factor.

| Variable | VIF | 1/VIF |
|---|---|---|
| FDI | 1.00 | 1.00 |
| GDP | 1.02 | 0.98 |
| DFI | 1.09 | 0.92 |
| TRD | 1.12 | 0.89 |
| INFL | 1.10 | 0.90 |
| Pol | 1.15 | 0.87 |
| Law | 1.14 | 0.88 |

Source: Author's estimation using STATA.

Table 2. Correlation matrix of the variables.

| Correlation | FDI | GDP | DFI | EDU | TRD | INFL | Pol | Law |
|---|---|---|---|---|---|---|---|---|
| FDI | 1 | | | | | | | |
| GDP | 0.0001 | 1 | | | | | | |
| DFI | 0.0131 | 0.0664 | 1 | | | | | |
| TRD | 0.2335 | 0.0113 | 0.2043 | 1 | | | | |
| INFL | 0.0179 | 0.0414 | 0.0163 | 0.6547 | 1 | | | |
| Pol | 0.1890 | 0.0359 | 0.3347 | 0.0683 | 0.1884 | 1 | 1 | |
| Law | 0.1118 | 0.1144 | 0.2209 | 0.0338 | 0.0300 | 0.4531 | 0.0222 | 1 |

Source: Derived from the author's calculations using STATA.

Table 3. Statistical description of the model's variables.

| Variable | Obs. | Mean | Standard deviation | Minimum | Maximum |
|---|---|---|---|---|---|
| FDI | 145 | 0.0530 | 0.0912 | −0.0373 | 0.7207 |
| GDP | 145 | 1015.386 | 1718.654 | 85.5457 | 8605.29 |
| DFI | 145 | 30.5175 | 16.9634 | 0.3182 | 74.6192 |
| TRD | 145 | 0.7786 | 0.0098 | −0.0031 | 1.073 |
| INFL | 145 | 0.0977 | 0.2068 | −0.6387 | 0.8765 |
| Pol | 145 | 0.3362 | 0.2635 | 0.0562 | 1.5548 |
| Law | 145 | 0.1552 | 0.1890 | 0.0001 | 0.9382 |

Source: Author's estimation using STATA.

corrected prior to regression analysis.

Inflation is notably variable, with a mean annual inflation of approximately 9.8% and a standard deviation of 20.7 percentage points. The minimum inflation is –63.87%, and the maximum is +87.65%, reflecting episodes of hyperinflation and subsequent deflation/stabilization. For example, Zimbabwe's well-documented hyperinflation in the 2000s would account for such outliers. Most country-years have more moderate inflation, with a median inflation rate around 5-10%. This high volatility in inflation could potentially deter investment due to unpredictability.

The Political Stability index has a mean of 0.336 with a standard deviation of 0.263, indicating that political stability is generally weak in the sample. This suggests investors in many of these countries face non-trivial political risks.

The Rule of Law index has a mean of approximately 0.155, with values ranging from near 0.155 to 0.938. This confirms that, on average, governance and legal institutions in the sample are weak, though there are improvements in some country-years.

These descriptive statistics highlight that countries with higher income and better institutional scores tend to have higher FDI, while countries experiencing extreme instability, such as Zimbabwe's hyperinflation period, see negative FDI flows and minimal DFI. This variation provides a useful backdrop for the regression analysis, which will formally test the influence of each factor on FDI.

**Empirical results of the model**

The relationship between DFIs and FDI was examined

**Table 4.** Fixed effect approach.

| FDI | Coef. | Standard Error | t | P > |t| | Sigma_u | Sigma_e | rho |
|---|---|---|---|---|---|---|---|
| GDP | 1.592076 | 0.1817356 | 8.76 | 0.000 | 1.3108299 | 0.97643107 | 0.64314099 |
| DFI | 0.0311491 | 0.0997546 | 0.31 | 0.755 | 1.3108299 | 0.97643107 | 0.64314099 |
| TRD | 0.830838 | 0.2345126 | 3.54 | 0.001 | 1.3108299 | 0.97643107 | 0.64314099 |
| INFL | 0.0009286 | 0.1351949 | 0.01 | 0.995 | 1.3108299 | 0.97643107 | 0.64314099 |
| Pol | −0.139431 | 0.5508704 | −0.25 | 0.801 | 1.3108299 | 0.97643107 | 0.64314099 |
| Law | 0.1653325 | 0.3746255 | 0.44 | 0.660 | 1.3108299 | 0.97643107 | 0.64314099 |
| Constant | −21.1658 | 3.82951 | −5.53 | 0.000 | 1.3108299 | 0.97643107 | 0.64314099 |

Source: Author's estimation using STATA.

**Table 5.** Random effect approach.

| FDI | Coef. | Standard Error | z | P > |z| | Sigma_u | Sigma_e | rho |
|---|---|---|---|---|---|---|---|
| GDP | 0.9627578 | 0.995459 | 9.67 | 0.000 | 0 | 0.97643107 | 0 |
| DFI | 0.1286091 | 0.1306798 | 0.98 | 0.325 | 0 | 0.97643107 | 0 |
| TRD | 0.5933398 | 0.280704 | 2.11 | 0.035 | 0 | 0.97643107 | 0 |
| INFL | 0.0510496 | 0.1339424 | 0.38 | 0.703 | 0 | 0.97643107 | 0 |
| Pol | 0.590256 | 0.6120377 | 0.96 | 0.335 | 0 | 0.97643107 | 0 |
| Law | 1.050129 | 0.4657569 | 2.25 | 0.024 | 0 | 0.97643107 | 0 |
| -Constant | −9.540069 | 2.38881 | −3.99 | 0.000 | 0 | 0.97643107 | 0 |

Source: Author's estimation using STATA.

**Table 6.** Hausman test.

| FDI | (b) Fixed | (B) Random | (b-B) Difference | Sqrt (diag(V_b-V_B)) S.E | Prob > Chi$^2$ | Chi$^2$(6) = (b-B)'[(V_b-V_B)^(−1)] (b-B) |
|---|---|---|---|---|---|---|
| GDP | 1.592076 | 0.9627578 | 0.6293177 | 0.1520475 | 0.000 | 47.24 |
| DFI | 0.0311491 | 0.12866091 | −0.09746 | 0.123538 | 0.000 | 47.24 |
| TRD | 0.830838 | 0.5933398 | 0.2374982 | 0.018359 | 0.000 | 47.24 |
| INFL | 0.0009286 | 0.0510496 | −0.050121 | 0.022210 | 0.000 | 47.24 |
| Pol | −0.139431 | 0.590256 | −0.7296871 | 0.133458 | 0.000 | 47.24 |
| Law | 0.1653325 | 1.050129 | −0.8847963 | 0.015762 | 0.000 | 47.24 |

b = Fixed-effects estimates (consistent under general conditions); B = random-effects estimates (efficient but potentially inconsistent). The test compares the two models. A significant result favors fixed-effects.
Source: Author's estimation using STATA.

using a panel data framework. To identify the most appropriate estimation technique, a Hausman specification test was conducted comparing the fixed effects (FE) and random effects (RE) models. This test assesses whether unobserved country-specific characteristics are correlated with the regressors, which would bias the RE estimator. Under the fixed effects model, all time-invariant factors specific to each country are absorbed into the entity-specific intercepts, allowing consistent estimation even when unobserved heterogeneity exists (Greene, 2012). In contrast, the random effects model assumes that these unobserved factors are uncorrelated with the independent variables, a less restrictive condition that may result in more efficient estimates if the assumption holds.

Although the fixed effects model is preferred, a random effects model was estimated for robustness. The direction and significance of the main variables remained consistent, validating the findings. The coefficients on GDP per capita and Trade openness remained strong and significant. The results of both models are presented in Tables 4 and 5, with each model estimated on a sample of 145 observations.

Once both estimation methods were applied, the saved output files were used to conduct the Hausman test, as summarized in Table 6. The Hausman test examines whether the model assumptions underlying the random-effects estimator hold. Specifically, it tests whether there is a systematic difference in the coefficients estimated by the

Table 7. Fixed effects generalized least squares regression 1990–2018.

| Variable | −1 Whole sample (SSA countries) | −2 Sectoral analysis (SSA countries) |
|---|---|---|
| GDP | 0.023 (−0.671) | 0.104 (−0.09) |
| DFI | 0.156 (−0.245) | 0.244 (−0.066) |
| TRD | 0.214 (−0.105) | 0.177 (−0.14) |
| INFL | −0.320 (0) | −0.154 (−0.003) |
| Pol | 0.186 (−0.001) | 0.022 (−0.607) |
| Law | 0.164 (−0.222) | 0.294 (−0.03) |
| DFI_AGRI | - | 0.055 (−0.366) |
| DFI_FIN | - | 0.237 (−0.085) |
| DFI_INFRA | - | 0.108 (−0.24) |
| Constant | 1.208 (−0.433) | 0.051 (−0.395) |
| Observations | 145 | 145 |
| $R^2$ | 0.52 | 0.618 |

Robust standard error errors are shown in parentheses, except for Wald Chi-square where the probability value is reported in parentheses.
Source: Author's estimation using STATA, accounting for both country and year fixed effects.

fixed- and random-effects models. A significant p-value supports the use of fixed effects (Hausman, 1978).

Based on the Hausman test, the fixed effects model was selected, as the null hypothesis favoring random effects was rejected (p < 0.001). This outcome implies that unobserved, time-invariant country characteristics are correlated with the regressors, making the fixed effects estimator more reliable for inference.

**Regression results**

As shown in Table 7, the panel regression yields an R-squared of 0.520, indicating that the independent variables explain approximately 52% of the variation in FDI inflows across the sample. All variables were log-transformed prior to estimation to allow interpretation of coefficients in terms of proportional change and to address distributional concerns (Wooldridge, 2010). This transformation improves interpretability: each coefficient now reflects the expected percentage change in FDI in response to a one-percent change in the corresponding independent variable.

The results show that GDP per capita has a positive and highly significant effect on FDI. Quantitatively, the result indicates that as GDP per capita rises (that is, as a country becomes wealthier or its market larger), FDI inflows increase, supporting H2. This aligns with prior studies emphasizing market size and economic strength as key drivers of FDI. In practical terms, foreign investors appear to favor the relatively more developed economies in SSA, likely due to better infrastructure, higher consumer purchasing power, and possibly more stable environments that come with development. This result is consistent with Abdi et al. (2024), who found that market size is a significant attractor of FDI in Africa.

The DFI variable is also significant, suggesting that multilateral finance initiatives play a positive role in catalyzing FDI, especially in environments where commercial investment might otherwise be constrained. This aligns with Massa (2011), who argues that DFIs help reduce perceived risk while supporting long-term growth by investing in sectors often underserved by private capital.

Trade openness is positive and statistically significant in the regression, confirming H4. A one-unit increase in the trade/GDP ratio (a substantial change) is associated with a considerable increase in FDI/GDP. More concretely, countries that are more open to trade tend to receive more FDI, all else equal. This makes intuitive sense: openness to trade often goes hand in hand with openness to investment, and many foreign investors seek export opportunities. This finding reinforces the notion that market access and integration into the global economy are important for attracting FDI.

Inflation (INFL), on the other hand, shows a negative association with FDI, which is in line with theoretical expectations (H3): rising prices may increase uncertainty and reduce investor confidence due to higher costs (Kore, 2019). According to Kore (2019), high inflation rates in an unstable economic environment typically increase the cost of investment, negatively affecting FDI returns.

The institutional variables, Political Stability and Rule of Law, produced mixed results. While the rule of law showed a positive and statistically significant relationship with FDI inflows (H6), political stability did not demonstrate a consistent effect. This may suggest that while legal certainty influences investment decisions, broader political factors may be more context-dependent or less impactful in some SSA countries. These findings align with prior

9Cagiza and Cagiza    9

Table 8. Sensitivity analysis and robustness check (whole sample 1990–2018).

| Variable | Model 1 | Model 2 | Model 3 | Model 4 | Model 5 | Model 6 | Model 7 | Model 8 |
|---|---|---|---|---|---|---|---|---|
| GDP | 0.025 (0.664) | 0.167 (0.249) | 0.026 (0.666) | 1.088 (0.000)** | 1.150 (0.000)** | 0.193 (0.552) | 0.362 (0.012)** | 0.011 (0.058) |
| DFI | 0.156 (0.243) | 0.198 (0.122) | 0.294 (0.036)** | 0.053 (0.465) | 0.063 (0.295) | 0.195 (0.133) | 1.296 (0.000)** | - |
| TRD | 0.219 (0.108) | 0.977 (0.000)*** | 0.978 (0.20) | 0.115 (0.345) | 0.252 (0.065) | 0.264 (0.055) | - | 0.039 (0.008)*** |
| INFL | −0.120 (0.021)** | −0.163 (0.002)** | −0.160 (0.005) | −0.343 (0.000)** | −0.287 (0.000)** | - | −0.139 (0.008) | 0.127 (0.022)*** |
| Pol | 1.582 (0.000)** | 0.109 (0.081) | 1.109 (0.017)** | 0.233 (0.120) | - | 1.482 (0.000)** | 0.396 (0.005)** | 0.078 (0.018)*** |
| Law | 0.188 (0.00)** | 0.240 (0.070) | 1.100 (0.000)** | - | 0.161 (0.244) | 0.098 (0.388) | 0.195 (0.150) | 0.036 (0.007)*** |
| Government effectiveness | 0.174 (0.189) | −0.157 (0.003) | - | −0.282 (0.000)** | −0.356 (0.000)** | −0.270 (0.000)** | 0.3624 (0.011)** | 0.014 (0.007)* |
| Control of corruption | 1.149 (0.000)** | - | 0.066 (0.267) | −0.423 (0.000)** | −0.308 (0.000)** | 0.095 (0.366) | −0.104 (0.561) | −0.0101 (0.062)* |
| Constant | 0.336 (0.074) | 0.026 (0.671) | 1.152 (0.000)** | 0.249 (0.068) | 0.043 (0.451) | 0.109 (0.29) | 0.099 (0.466) | 0.048 (0.037) |
| Observations | 145 | 145 | 145 | 145 | 145 | 145 | 145 | 145 |
| R2 | 0.42 | 0.46 | 0.50 | 0.40 | 0.55 | 0.60 | 0.48 | 0.59 |

Standard errors in parentheses; ***p < 0.01, **p < 0.05, *p < 0.1.
Source: Author's estimation using STATA (country and year effects).

research indicating that improvements in governance and legal frameworks can promote investor confidence (Kore, 2019; Sabir et al., 2019).

The study's findings carry important implications for policymakers across SSA. In particular, the analysis suggests that in environments with weak legal systems or limited institutional capacity, attracting FDI becomes more challenging. Strengthening institutional frameworks may be necessary not just for transparency and fairness but also to support macroeconomic objectives. These results reinforce the argument that sustainable FDI attraction requires legal and institutional credibility, not just market openness or growth potential.

### Sectoral analysis

At the sectoral level, Table 7 shows that infrastructure (INFRA) investments supported by DFIs had the strongest positive influence on FDI inflows, with a coefficient of 0.108 (p = 0.24). Agribusiness (AGRI) followed, showing a moderate association with FDI. Financial sector interventions (FIN) were also positively related to FDI, with a coefficient of 0.237 (p = 0.085), indicating that targeted DFI involvement in banking and financial inclusion may facilitate investment under the right conditions. These results are consistent with Massa's (2011) observation that DFIs often target critical economic sectors where private capital is scarce. By funding these sectors, DFIs help reduce perceived risks and unlock investment opportunities. In the context of the SDGs, the findings suggest that public policy, especially those reinforcing institutional strength, can play a critical role in directing private investment toward high-impact sectors and supporting long-term development goals.

### Robustness

To evaluate the consistency of the findings, a series of robustness checks were conducted by alternately including and excluding selected independent variables from the regression models. Specifically, eight separate fixed effects regressions were run using FDI as the outcome variable (Table 8). This approach helped identify whether core variable relationships were stable across model specifications.

In addition to the main predictors, two governance-related indicators, Government Effectiveness and Control of Corruption, were incorporated based on their relevance in the literature as institutional drivers of FDI. These supplementary variables were sourced from the World Bank and matched to the structure of the primary dataset to ensure comparability.

The results from the sensitivity analysis indicate that changes in model specification, such as including or excluding certain predictors, do not materially affect the direction or significance of the key variables, including DFIs. This stability enhances confidence in the reliability of the findings.

**CONCLUSION, LIMITATION AND RECOMMENDATION**

This study examined the relationship between DFIs, FDI, and economic growth in SSA, focusing on whether DFIs help catalyze FDI inflows and under what conditions. Using panel data from 1990 to 2018 for five SSA countries, a framework was tested wherein DFIs, macroeconomic indicators (GDP per capita, trade openness, inflation), and institutional quality (political stability, rule of law) jointly influence FDI. A sectoral analysis of DFI commitments (in infrastructure, finance, and agribusiness) was also performed to pinpoint where DFIs might have the greatest impact on attracting FDI.

The findings provide nuanced insights into the role of DFIs in promoting FDI and economic growth in SSA. It shows that GDP per capita, trade openness, and rule of law are robust and significant predictors of FDI in the region. These results affirm the importance of economic size, liberal trade regimes, and strong institutions in attracting international investment. Contrary to expectations, total DFI commitments were not statistically significant in determining FDI inflows when controlling for these other factors. However, this does not negate the theoretical importance of the relationship. Instead, it highlights the complex interplay between DFIs and other economic and institutional factors.

When disaggregated by sector, DFIs directed toward infrastructure and, to a lesser extent, financial services, exhibited a positive and meaningful association with FDI. These results imply that DFIs may exert a more sector-specific and conditional influence on foreign investment, particularly when aligned with enabling sectors and broader reform efforts. In terms of institutional quality, the results indicate that while the rule of law significantly enhances FDI, political stability did not have a similar effect. This could be attributed to political dynamics within the selected SSA countries. This discrepancy suggests that while legal frameworks are important, the stability of political environments might not be uniformly beneficial in all contexts.

By systematically examining the macro-level dynamics between DFIs and FDI, this study fills a notable gap in the literature. While previous research has often focused on either the micro-level impact of DFIs or the determinants of FDI in developing economies, this work integrates both dimensions, offering a region-specific understanding of their interaction in SSA. Beyond empirical contributions, this study advances the development finance discourse by offering longitudinal evidence on the evolving role of DFIs in SSA. It also highlights the increasing relevance of private-sector-led financing mechanisms compared to traditional aid. The mixed findings on inflation and FDI signal opportunities for further research into the specific economic environments of SSA countries, with potential implications for refining economic theory. From a policy perspective, the results suggest a few key takeaways:

1. Leverage DFIs in infrastructure: Governments and international partners might focus DFI resources on infrastructure projects, as these have the strongest pull-through effect on FDI. Ensuring that DFI-funded infrastructure is well-planned and aligns with areas of interest to foreign investors could maximize the impact.
2. Strengthen legal institutions: Reforms that strengthen the rule of law, such as judicial reforms, anti-corruption measures, and property rights protections, could significantly increase FDI inflows. The findings underscore rule of law as a more consistent determinant of FDI than short-term political stability, suggesting that maintaining a strong legal framework can mitigate investor fears.
3. Maintain openness and market growth: Strategies to sustain economic growth and keep trade open, including low trade barriers and participation in regional trade agreements, remain fundamental. These factors directly contribute to making a country attractive for FDI and likely create a virtuous cycle where more FDI fuels further growth and trade integration.

This study has several limitations that merit consideration:

1. Data limitations: The dataset ends in 2018, and more recent developments, such as COVID-19 impacts, are not captured. The unavailability of disaggregated DFI data post-2018 constrained the analysis.
2. Endogeneity: The model does not fully account for reverse causality, where countries receiving more FDI may also attract more DFIs. Future studies should explore instrumental variable or dynamic panel techniques to mitigate this concern.
3. Country coverage: The study is limited to five SSA countries due to data availability and comparability constraints. While these countries are diverse, broader generalization should be approached cautiously.

Despite these limitations, the study provides a useful empirical baseline on the interplay of DFIs, FDI, and domestic conditions in SSA.

Future research could broaden the sample size, incorporate additional explanatory variables (e.g., environmental indicators or corruption indices), apply dynamic panel models to account for feedback effects over time, and explore comparative analyses with other developing regions. Policy simulation approaches could also be employed to model potential institutional reforms and their impact on FDI and growth.

This study contributes meaningfully to both academic literature and policy debates by clarifying the role of DFIs in fostering economic development in SSA. By linking DFIs, FDI, institutional quality, and sectoral priorities, it provides a comprehensive framework for future research and policy action toward more effective use of development finance in promoting sustainable growth across the region.

**CONFLICT OF INTERESTS**

The authors have not declared any conflict of interests.

## ACKNOWLEDGEMENTS

The authors thank the University of Southern Mississippi, The World Bank Group, and the Permanent Mission of Angola to the United Nations. This article is based in part on the first author's PhD dissertation completed at the University of Southern Mississippi. The authors acknowledge the guidance and academic environment that made the original research possible.